# Energetics, Charge Transfer and Magnetism of Small Molecules Physisorbed on Phosphorene

Yongqing Cai[†], Qingqing Ke[‡], Gang Zhang[*,†], and Yong-Wei Zhang[*,†]

[†]Institute of High Performance Computing, A*STAR, Singapore 138632

[‡]Department of Materials Science and Engineering, National University of Singapore, Singapore 117574

**ABSTRACT**: First-principles calculations are performed to investigate the interaction of physisorbed small molecules, including CO, $H_2$, $H_2O$, $NH_3$, NO, $NO_2$, and $O_2$, with phosphorene, and their energetics, charge transfer, and magnetic moment are evaluated on the basis of dispersion corrected density functional theory. Our calculations reveal that CO, $H_2$, $H_2O$ and $NH_3$ molecules act as a weak donor, whereas $O_2$ and $NO_2$ act as a strong acceptor. While NO molecule donates electrons to graphene, it receives electrons from phosphorene. Among all the investigated molecules, $NO_2$ has the strongest interaction through hybridizing its frontier orbitals with the 3p orbital of phosphorene. The nontrivial and distinct charge transfer occurring between phosphorene and these physisorbed molecules not only renders phosphorene promising for application as a gas sensor, but also provides an effective route to modulating the polarity and density of carriers in phosphorene. In addition, the binding energy of $H_2$ on phosphorene is found to be 0.13 eV/$H_2$, indicating that phosphorene is suitable for both stable room-temperature hydrogen storage and its subsequent facile release.

*Supporting Information Placeholder*

## INTRODUCTION

Recent years have witnessed rapid progress in the synthesis, characterization, and applications of atomically thin two-dimensional (2D) materials, such as graphene and transition metal dichalcogenide (TMD).[1,2] Graphene has many fascinating properties, based on which, many proof-of-concept devices, such as transistors, supercapacitors and gas sensors, have been demonstrated.[3-6] Nanostructured $MoS_2$, with a size-independent carrier mobility[7] and a much smaller thermal conductivity than graphene,[8,9] has also attracted increasing attention for applications in nanoelectronic and thermoelectric.[10,11] On the one hand, owing to their ultrathin thickness, high surface-volume ratio and weak electronic screening, properties of monolayer graphene and $MoS_2$ tend to be strongly affected by physical/chemical adsorbates.[12-14] On the other hand, a promoted interaction between external molecules and 2D materials not only renders them eligible for sensing applications but also enables the modulation of their electronic and chemical properties.[15-17] For instance, surface modification via hydrogenation was shown to induce a band gap opening in graphene,[18] and ordered hydrogen adsorption on $MoS_2$ was shown to create a conducting nanoroad.[19]

Small molecules, such as CO, $H_2$, $H_2O$, $NH_3$, NO, $NO_2$, and $O_2$, are known to be ubiquitously present on 2D materials surfaces, and unsurprisingly, these small molecules can never be fully removed from these surfaces due to their large surface areas.[11] Importantly, physisorbed small molecules were found to create marked effects in modifying carriers density, inducing the shift of Fermi level, and modulating the optical properties of graphene and $MoS_2$.[20-27] In fact, due to these effects, graphene and $MoS_2$ were shown to be used as a gas sensor.[13-16]

Recently, phosphorene,[28] a new elemental 2D material, has been exfoliated by mechanical cleavage of black phosphorus (BP), the most stable allotrope of the element phosphorus in ambient condition.[29] A sizable direct band gap (1.5 eV for monolayer phosphorene) and a large room-temperature mobility up to 1000 $cm^2V^{-1}s^{-1}$ render phosphorene promising for applications in nanoelectronics and optoelectronics.[30-32] Different from graphene and $MoS_2$, phosphorene is an intrinsic p-type semiconductor. It is noted that many studies have been devoted to examining various factors on the electronic properties of phosphorene. For example, strong thickness-dependent electronic properties of few-layer phosphorene such as work function, band gap, band alignment have been examined.[33] The effects of layer disorder[34] and nanoribbons width,[35] and also electronic anisotropy[36-40] of phosphorene have also been analyzed. Despite of these interesting studies, systematic understanding of the effect of physisorbed small molecules, such as CO, $H_2$, $H_2O$, $NH_3$, NO, $NO_2$, and $O_2$, on the electronic properties of phosphorene is still lacking. Owing to the marked effects of physisorbed small molecules on graphene and $MoS_2$, clearly, it is both of scientific interest and technological importance to understand the energetics, charge transfer and magnetism of small molecules physisorbed on phosphorene.

In this work, we perform a systematic theoretical study on the adsorption of small molecules (CO, $H_2$, $H_2O$, $NH_3$, NO, $NO_2$ and $O_2$) on phosphorene, focusing on the energetics, magnetic moments, and charge transfer between these molecules and phosphorene. We find that CO, $H_2$, $H_2O$, and $NH_3$ act as charge donors; whereas $NO_2$, and $O_2$ serve as charge acceptors. While NO molecule donates electrons to graphene, it accepts electrons from phosphorene. Our present study provides a valuable reference for assessing the effect of physisorbed small molecules on phosphorene. In addition, the present work also demonstrates that the electronic properties of phosphorene can be tuned by selective adsorption of those small molecules.

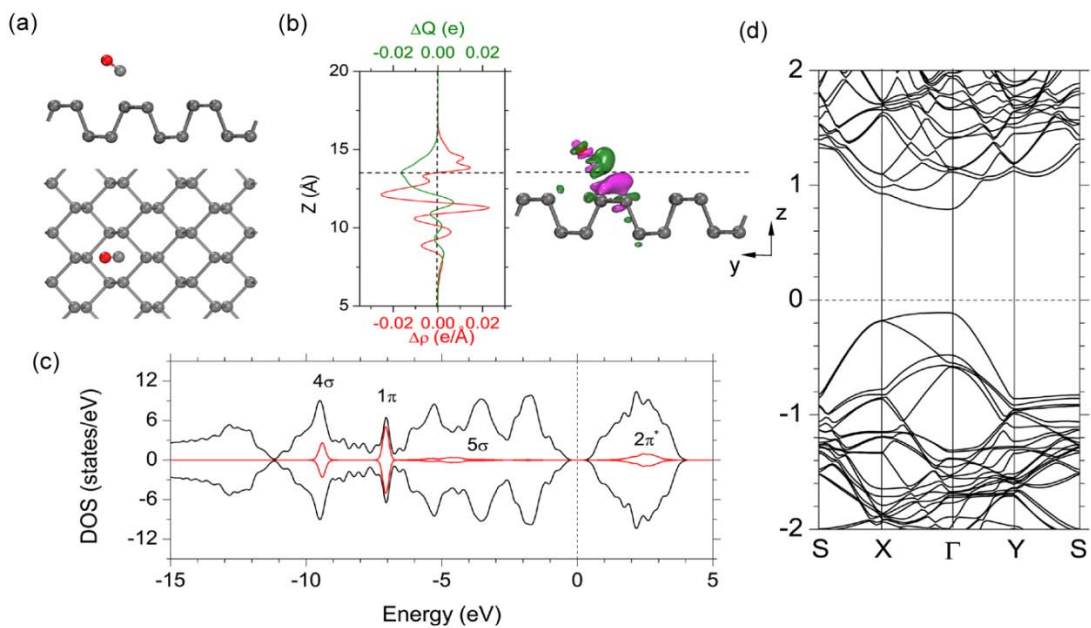

Fig. 1 CO adsorbed on phosphorene. (a) The side and top views of the lowest-energy configuration. The phosphorus, carbon and oxygen atoms are represented by balls in grey, black and red colors, respectively. (b) Plane-averaged differential charge density $\Delta\rho(z)$ (red line), amount of transferred charge $\Delta Q(z)$ (green line), and side view of the 0.005 Å$^{-3}$ DCD isosurface (right). The violet (green) color denotes diminishing (accumulation) of electrons. (c) DOS of CO on phosphorene (black line), LDOS and positions of the molecular orbitals of CO (red line). (d) Band structure of CO on phosphorene.

## COMPUTATIONAL METHODS

The first-principles calculations are performed within the framework of density functional theory (DFT) by using Vienna ab initio simulation package (VASP) package.[41] Van der Waals (vdW) corrected functional with Becke88 optimization (optB88)[42] is adopted for analyzing the noncovalent chemical functionalization of phosphorene by small molecules. All the structures are relaxed until the forces exerted on each atom are less than 0.005 eV/Å. The relaxed lattice constant of monolayer phosphorene is $a$=3.335 Å, $b$=4.571 Å along zigzag and armchair directions, respectively based on a 14×10×1 Monkhorst-Pack (MP) grid for k-point sampling. Here we only consider a single molecule adsorption in the 4×3×1 supercell (48 phosphorus atoms) in the dilute doping limit. The thickness of the vacuum region is greater than 15 Å. The first Brillouin zone is sampled with a 3×3×1 MP grid and a kinetic energy cutoff of 400 eV is adopted. The binding energy ($E_b$) of the molecule to phosphorene is calculated as $E_{Mol+P}-E_{Mol}-E_P$, where $E_{Mol}$, $E_P$ and $E_{Mol+P}$ are the energies of the molecule, phosphorene sheet and molecule adsorbed phosphorene, respectively.

## RESULTS

We have examined several possible anchoring positions on the high symmetric sites of all the molecules on the phosphorene surface, including both above the puckered hexagon and the zigzag trough with the molecules being aligned either parallel or perpendicular to the surface, and the results based on the lowest-energy configurations are compiled in Table 1.

### Molecular Donors

In the following, we examine four molecules: CO, $H_2$, $H_2O$, and $NH_3$, which have a closed-shell electronic structure and serve as donors.

**CO adsorption:** For CO adsorption, the most stable binding configuration is shown in Fig. 1a, where the molecule locates above the ridge with the CO bond being aligned perpendicular to the surface with a slight deviation from the normal direction. The C atom has a coordination number of four and forms three weak vdW C-P bonds with the bond length ranges from 3.06 to 3.15 Å. The vdW-corrected $E_b$ for single CO is -0.31 eV.

To analyze the electronic interaction between the molecule and phosphorene, we calculate the differential charge density (DCD) $\Delta\rho(r)$. The isosurface of the $\Delta\rho(r)$ for the adsorbed CO molecule is depicted in Fig. 1b. It can be seen that there is a large charge redistribution upon CO adsorption, especially between the adsorption gap of the molecules and the P atoms, which significantly disturbs the HOMO state (5σ), which is mainly located on the C atom (see the inset of Fig. 1c). This is also clearly reflected by the local density of states (LDOS) analysis as shown in Fig. 1c, where the CO 5σ state is widely broadened and distributed from -6 to -4 eV below the Fermi level; while the CO 2π*(LUMO), 1π (HOMO-1) and 4σ (HOMO-2) states are less affected after the adsorption. For pristine phosphorene, the three peaks below the valence band edge, which are located at around -2, -3.5, -5 eV, consist of mainly ($p_y$,$p_z$), ($p_x$,$p_y$,$p_z$), ($p_x$,$p_y$,$p_z$) hybridizations with a small amount of 3s orbital. The broad peaks at -10 eV and -13 eV are mainly comprised of 3s state. This indicates that the strong broadening of CO 5σ state mainly arises from the interaction with the phosphorus 3p orbitals.

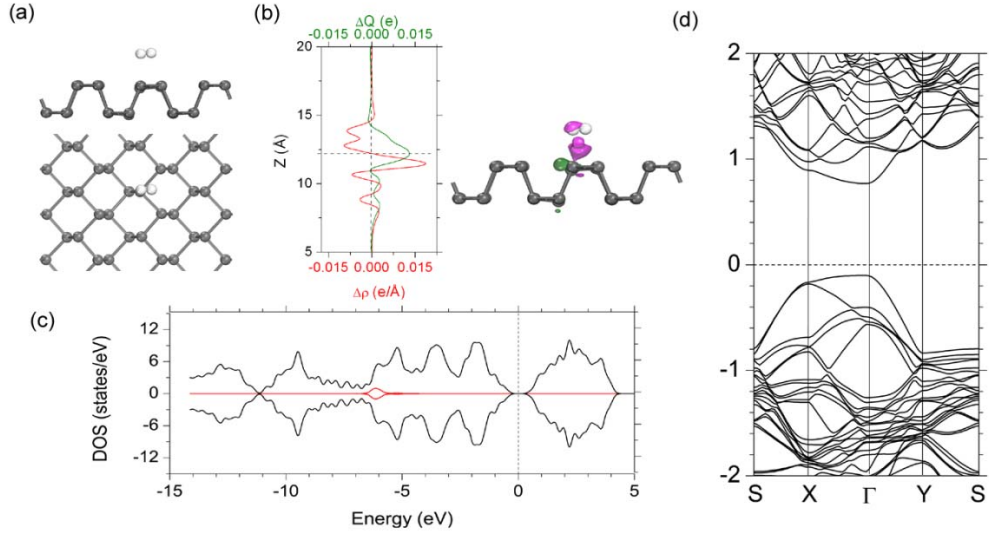

Fig. 2 $H_2$ adsorbed on phosphorene. (a) The side and top views of the lowest-energy configuration. The phosphorus and hydrogen atoms are represented by balls in grey and white colors, respectively. (b) Plane-averaged differential charge density $\Delta\rho(z)$ (red line), amount of transferred charge $\Delta Q(z)$ (green line), and side view of the 0.005 Å$^{-3}$ DCD isosurface (right). The violet (green) color denotes diminishing (accumulation) of electrons. (c) DOS of $H_2$ on phosphorene (black line) and LDOS of $H_2$ (red line). (d) Band structure of $H_2$ on phosphorene.

To quantify the amount of charge transfer from the molecule, we calculate the plane-averaged DCD $\Delta\rho(z)$ along the direction normal to the surface by integrating $\Delta\rho(r)$ within the x-y plane. The amount of transferred charge up to $z$ point is obtained using $\Delta Q(z) = \int_{-\infty}^{z} \Delta\rho(z')\,dz'$. We plot the $\Delta\rho(z)$ and $\Delta Q(z)$ together with the DCD in Fig. 1b for comparison. The integral of the charge accumulated from the bottom vacuum up to the interface between the adsorbate and the phosphorene (defined as the plane of zero charge variation shown in the $\Delta\rho(z)$ curve) gives the total charge donated by the molecule to the phosphorene. The exact amount of charge transfer normally corresponds to the maximum of the $\Delta Q(z)$ curve at the interface of the adsorbate and the surface.[17] It can be seen that the total charge transfer from the CO to phosphorene amounts to 0.007 $e$. Therefore, the CO molecule acts as a weak donor on the phosphorene surface, and the bands of phosphorene (Fig. 1d) are almost unchanged upon CO adsorption. The C-O bond length of the adsorbed CO is 1.15 Å, which is only slightly increased from that (1.14 Å) of its gas molecule. Our results are in good agreement with a recent work using local density approximation (LDA) approach in terms of the adsorption configuration and charge transfer between CO and phosphorene.[43]

**$H_2$ adsorption:** The most stable configuration for $H_2$ adsorption adopts a nearly parallel configuration with the H-H bond being along the armchair direction, and one of the H atom being directly above the P atom with the H-P bond length of 2.46 Å (Fig. 2a). The $E_b$ is found to be -0.13 eV. The value of $E_b$ is found to be -0.13 eV, which is close to the requirement for an appropriate hydrogen storage with a binding energy of -0.15– -0.3 eV/$H_2$.[44,45] The intermediate binding energy of hydrogen molecule on phosphorene implies a stable hydrogen storage at ambient conditions and subsequent facile release. It is noted that the $E_b$ of $H_2$ on phosphorene is much larger than that of $H_2$ on pristine graphene (~ 0.04 eV/$H_2$),[46] which suggests a higher working temperature for using phosphorene as the hydrogen adsorbent. However, since the density of bulk phosphorene is about 23% larger than that of graphite, implying a smaller gravimetric value of the hydrogen storage than graphite. Nevertheless, the highly flexible and puckered honeycomb structure of phosphorene implies a higher possibility for increasing the chemical affinity for hydrogen with convex areas of phosphorene surfaces, which has been demonstrated in fullerenes, nanotubes and graphene by controlling curvature or creating ripples.[47-51]

The isosurface plot of DCD (Fig. 2b) shows that there is a depletion of electrons in H atoms and an accumulation of electrons in the nearest P atom. The hydrogen molecule clearly donates electrons to phosphorene with around 0.013 $e$ per molecule. Compared with CO molecule, while the $H_2$ molecule has a larger amount of charge transfer, its binding energy is smaller. The underlying reason is that the binding mechanism of molecules on surface can be factored into two groups: the covalent interactions accompanying with the charge transfer and electrostatic attractions associated with polarization induced dipole-dipole interaction. For physisorption, the latter can play a more important role. Since the dipole moment of $H_2$ is zero, the dispersive force and polar-polar interaction between $H_2$ and phosphorene are much weaker than that of the strongly polarized CO molecule, leading to the lower binding energy despite the fact that the $H_2$ has a larger charge transfer. The relatively weak interaction is also reflected by the sharp peak in the LDOS of $H_2$ and the nearly unchanged band structure of adsorbed phosphorene compared with that of the pristine phosphorene, as shown in Fig. 2c and d, respectively.

**$H_2O$ adsorption:** For the $H_2O$ molecule, we have considered various orientations including the two O-H bonds pointing up or down or parallel to the surface. The most stable configuration is given in Fig. 3a, where one of the O-H bonds is parallel to the surface along the armchair direction and the other nearly normal to the surface. The in-plane O-H bond locates directly above the armchair P-P bond with the O-P bond length of 2.71 Å, and two H-P bonds of 3.14 and 3.21 Å. The binding energy is -0.14 eV, almost same as $H_2$ adsorption. Notably, this configuration is quite similar to the structure of $H_2O$ molecule adsorbed on graphite surface.[52]

The charge transfer analysis shows that each $H_2O$ donates about 0.035 $e$. The relatively large dipole moment of $H_2O$ induces a large charge redistribution even in the opposite side of the adsorbed phosphorene surface.(Fig. 3b) This prominent effect of water on the electronic properties of phosphorene shows that in real device, the water may significantly affect the performance such as durability and carriers mobility. The three highest occupied orbitals of the water molecule are $1b_1$ (HOMO), $3a_1$ (HOMO-1), and $1b_2$ (HOMO-2), which are orthogonal around the oxygen atom. The $3a_1$ state with the orbital nearly being normal to the surface has the largest orbital mixing with the P atom, which allows more efficient charge transfer to the phosphorene layer. This interaction in turn makes the energy level of $3a_1$ state greatly broadened (Fig. 3c) and also a slightly downward shift of this state from the HOMO level compared with the energy diagram of the isolated molecule. Similar to CO and $H_2$, there is no localized states within the band gap of $H_2O$ decorated phosphorene according to Fig. 3d.

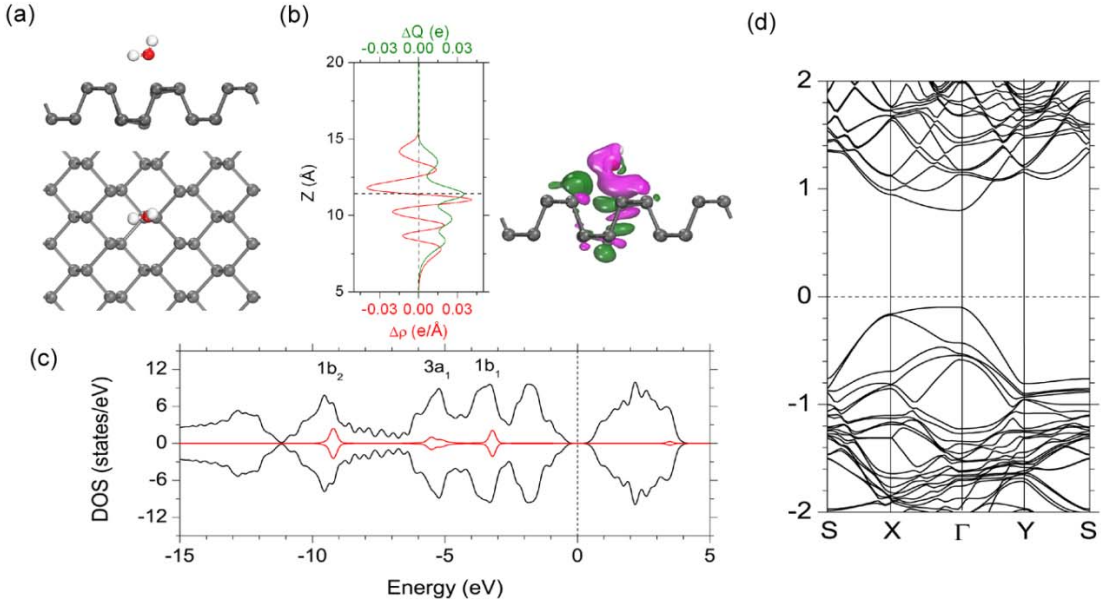

Fig. 3 $H_2O$ adsorbed on phosphorene. (a) The side and top views of the lowest-energy configuration. The phosphorus, oxygen and hydrogen atoms are represented by balls in grey, red and white colors, respectively. (b) Plane-averaged differential charge density $\Delta\rho(z)$ (red line), amount of transferred charge $\Delta Q(z)$ (green line), and side view of the 0.005 Å$^{-3}$ DCD isosurface (right). The violet (green) color denotes diminishing (accumulation) of electrons. (c) DOS of $H_2O$ on phosphorene (black line) and LDOS and positions of the molecular orbitals of $H_2O$ (red line). (d) Band structure of $H_2O$ on phosphorene.

Interestingly, while phosphorene has the same topologically equivalent honeycomb lattice as graphene, the charge state of $H_2O$ is quite different for adsorptions on phosphorene and graphene. Here we show that $H_2O$ acts as a strong donor on monolayer phosphorene surface, in contrast, previous studies show that $H_2O$ is a charge acceptor on graphene surface.[16] This distinctive behavior is attributed to the much larger work function of phosphorene (5.2 eV for monolayer phosphorene), thus more electron transferred from water to phosphorene, compared with that of graphene (4.5 eV). Since the phosphorene layer is an intrinsically p-type material with a wealth of holes located at the valence band, the decrease of work function and the upward shift of valence band with increasing the thickness of phosphorene reported recently[33] suggest that the charge transfer from water molecule tends to be highly sensitive to the thickness of phosphorene with less electrons being donated to the multilayer phosphorene compared with the monolayer phosphorene.

**NH$_3$ adsorption:** Two different adsorption orientations are considered for NH$_3$ molecule: the H atoms pointing away from the surface, and the H atoms pointing toward the surface. The strongest binding site is found to be the former with the N atom located directly above the P atom (Fig.4a). In this binding scenario, the N-P bond length is 2.59 Å and the binding energy is -0.18 eV. As depicted in Fig. 4b, the DCD analysis shows that there is a significant charge redistribution between the NH$_3$ and phosphorene, indicating moderate covalent bonding between NH$_3$ and phosphorene surface arising mainly from the ionization of the lone electron pair at the N atom in the $3a_1$ (HOMO) state. The NH$_3$ molecule acts as a donor by transferring around 0.05 $e$ to phosphorene. The nonbonding $3a_1$ orbital of NH$_3$ tends to be more affected by the P atoms as evidenced by the strongly broadened LDOS at the top of the valence band as shown in Fig. 4c. Regarding the band structure, the strong interaction and charge donation alter the top valence band of phosphorene along the Y-S and S-X directions with an upward shift of the band due to the enhanced coulomb interaction.

**Molecular Acceptors**

We next consider three paramagnetic molecules: NO, NO$_2$, and O$_2$, which have an open-shell electronic structure.

**NO adsorption:** For NO molecule, we consider the same initial adsorption configurations as CO molecule. The lowest energy state (Fig. 5a) is the same as CO molecule, with the N atom having a coordination number of four and forming three N-P bonds with bond length ranging from 2.75-2.88 Å, which are smaller than those of CO molecule. The binding energy is found to be -0.32 eV, which is almost the same as that of CO molecule.

The half-filled doubly degenerated NO frontier orbital $2\pi$ (HOMO), together with its spin-split partner (LUMO), allows back and forward charge transfer by orbital hybridization with phosphorus orbitals. As shown in Fig. 5b, the plane-averaged DCD shows that there is a charge accumulation in NO molecule whereas a loss of electron of phosphorene. The total amount of transferred charge from phosphorene is 0.074 $e$. According to the band structure (Fig. 5d), the singly occupied HOMO state of NO moves above the Fermi level and becomes depopulated after contacting with phosphorene. Therefore, the net gain of electrons from phosphorene results from a strong orbital mixing of HOMO-1 and HOMO-2 states with the 3s states of phosphorus, as reflected by the LDOS analysis as shown in Fig. 5c. This orbital mixing leads to a larger charge transfer induced by the depopulation of the HOMO state, and simultaneously a magnetic moment of 0.63 $\mu_B$ of the adsorbed NO. It is noted that the acceptor role of NO on phosphene is different from that of NO on graphene,[16] in which NO always donates electrons to graphene. It should be noted that a recent work by Kou et al.[43] predicts a different lowest-energy configuration, with NO binding directly above P atom. This difference could be attributed to the different exchange-correlation functional (LDA) and the vdW corrections (DFT-D2) used in their work. Nevertheless, the oxidation state of NO molecule on phosphorene predicted by both methods is the same. Moreover, our additional calculations show that the "acceptor" character of NO molecule is independent of the anchoring sites on the surface.

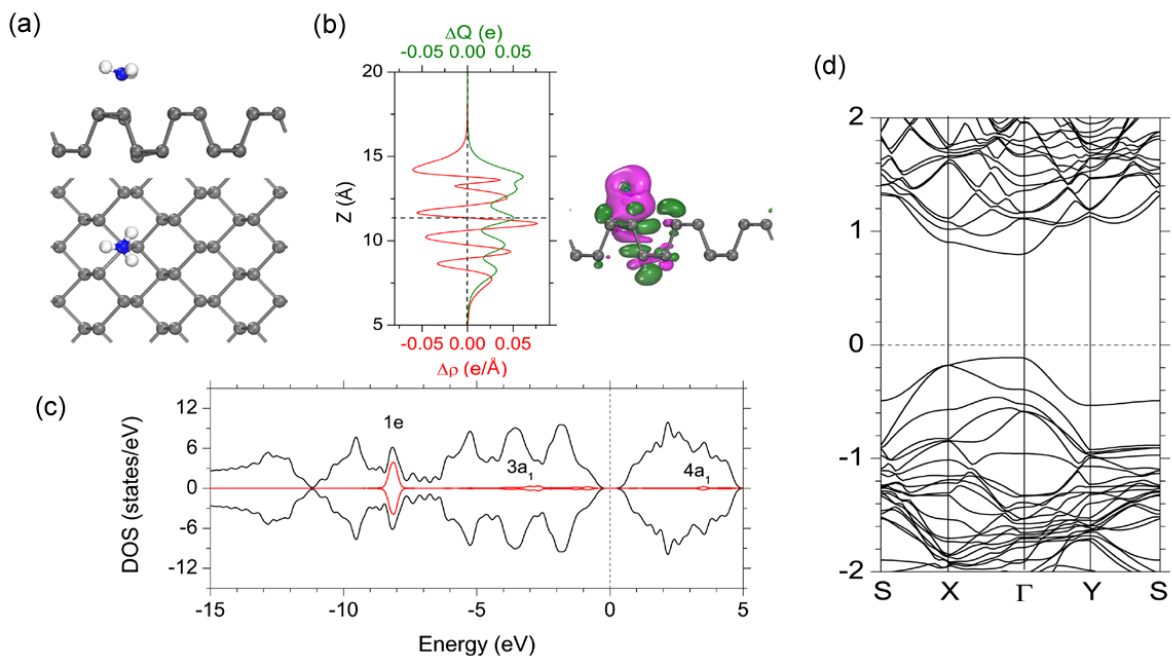

Fig. 4 NH₃ adsorbed on phosphorene. (a) The side and top views of the lowest-energy configuration. The phosphorus, nitrogen and hydrogen atoms are represented by balls in grey, blue and white colors, respectively. (b) Plane-averaged differential charge density $\Delta\rho(z)$ (red line), amount of transferred charge $\Delta Q(z)$ (green line), and side view of the 0.005 Å$^{-3}$ DCD isosurface (right). The violet (green) color denotes diminishing (accumulation) of electrons. (c) DOS of NH₃ on phosphorene (black line) and LDOS and positions of the molecular orbitals of NH₃ (red line). (d) Band structure of NH₃ on phosphorene.

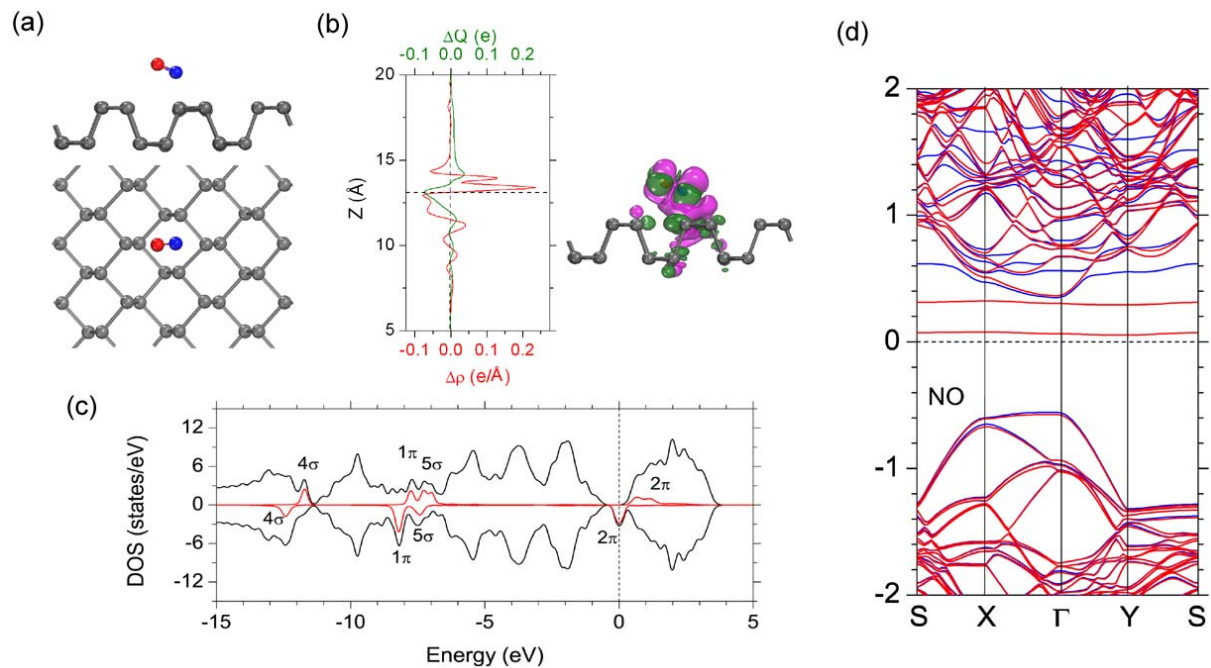

Fig. 5 NO adsorbed on phosphorene. (a) The side and top views of the lowest-energy configuration. The phosphorus, oxygen and nitrogen atoms are represented by balls in grey, red and blue colors, respectively. (b) Plane-averaged differential charge density $\Delta\rho(z)$ (red line), amount of transferred charge $\Delta Q(z)$ (green line), and side view of the 0.005 Å$^{-3}$ DCD isosurface (right). The violet (green) color denotes diminishing (accumulation) of electrons. (c) DOS of NO on phosphorene (black line) and LDOS and positions of the molecular orbitals of NO (red line). (d) Spin-polarized band structures of NO on phosphorene with spin-up bands and spin-down bands are indicated in blue and red, respectively.

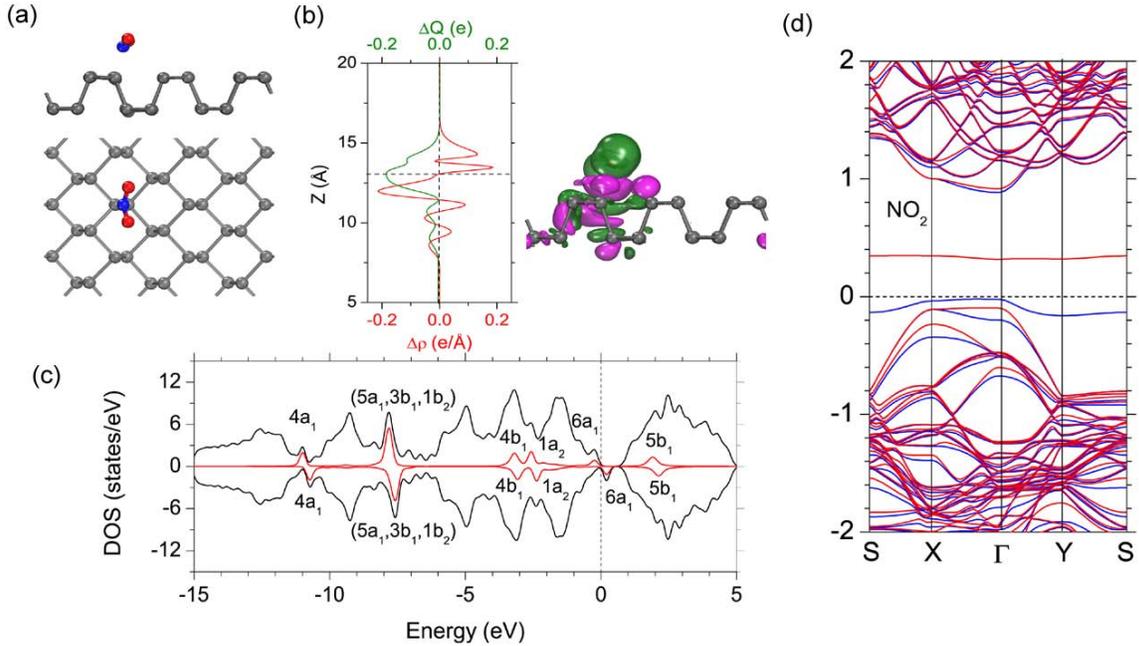

Fig. 6 NO$_2$ adsorbed on phosphorene. (a) The side and top views of the lowest-energy configuration. The phosphorus, oxygen and nitrogen atoms are represented by balls in grey, red and blue colors, respectively. (b) Plane-averaged differential charge density $\Delta\rho(z)$ (red line), amount of transferred charge $\Delta Q(z)$ (green line), and side view of the 0.005 Å$^{-3}$ DCD isosurface (right). The violet (green) color denotes diminishing (accumulation) of electrons. (c) DOS of NO$_2$ on phosphorene (black line) and LDOS and positions of the molecular orbitals of NO$_2$ (red line). (d) Spin-polarized band structures of NO$_2$ on phosphorene with spin-up bands and spin-down bands are indicated in blue and red, respectively.

**NO$_2$ adsorption:** We have placed a single NO$_2$ molecule on the surface with different orientations: the N-O bonds are oriented up, down or parallel to the surface and aligned along the armchair or zigzag directions. The configuration with the lowest energy is plotted in Fig. 6a. Different from other molecules, the most stable configuration of adsorbed NO$_2$ is found to be aligned parallel to the zigzag direction with the N atom directly above the P atom and the N-P bond length of 2.27 Å, which is different from NO$_2$ adsorption on graphene, where the bonding occurs through both oxygen atoms.[20] The NO$_2$ has the largest binding energy of -0.50 eV among all the studied molecules in present work. It acts as a strong acceptor with a charge transfer from phosphorene to NO$_2$ of 0.19 e based on DCD analysis given in Fig. 6b. It should be noted that our predicted lowest-energy configuration of NO$_2$ is different from the recent work[43] where the LDA and DFT-D2 were used. Nevertheless, the results of the oxidation state of the bound NO$_2$ molecule in both works are consistent. According to Fig.6c and d, the HOMO (6a$_1$,up) is slightly below E$_f$ and the LUMO (6a$_1$,down) is around 0.2 eV above the Fermi energy level E$_f$. Hence, NO$_2$ accepts electrons from phosphorene. Similar to NO, charge transfer from phosphorene to NO$_2$ occurs mainly through orbital hybridization, which simultaneously induces about 0.14 $\mu_B$ magnetic moment in phosphorene. This shows that the magnetic properties of phosphorene can be altered through a proper molecular functionalization.

**O$_2$ adsorption:** For the oxygen molecule, the most stable configuration is plotted in Fig. 7a, where the O-O bond is aligned along the armchair direction and forms an angle of around 30° with the surface. One of the O atoms is located above the P atom with a O-P bond length of 2.76 Å, and the binding energy is -0.27 eV. It should be noted that the O$_2$ molecule adopts a similar adsorption configuration as H$_2$ with one H or O atom being located right above the P atom, while for CO and NO, the C or N atom is located in the center of the puckered hexagon with a coordination number of four. The underlying reason may be due to the difference in dispersive interactions associated with the non-polar H$_2$ and O$_2$ molecules and the polar CO and NO molecules. Similar to NO and NO$_2$, O$_2$ molecule acts as a charge acceptor by gaining about 0.06 e from phosphorene, according to Fig. 7b. Clearly, O$_2$ has a weak physisorption on phosphorene and the O-O bond increases slightly from 1.22 Å of the free gas molecule to 1.24 Å for the bound molecule. The orbitals of O$_2$ is less affected by phosphorene despite a slight broadening of HOMO state (2π, up) (Fig. 7c). The antibonding LUMO state (2π, down) is nearly unaffected and located in the band gap of phosphorene (Fig. 7d). The magnetic moment of the adsorbed O$_2$ is around 1.54 $\mu_B$, which is slightly quenched from that of the free molecule.

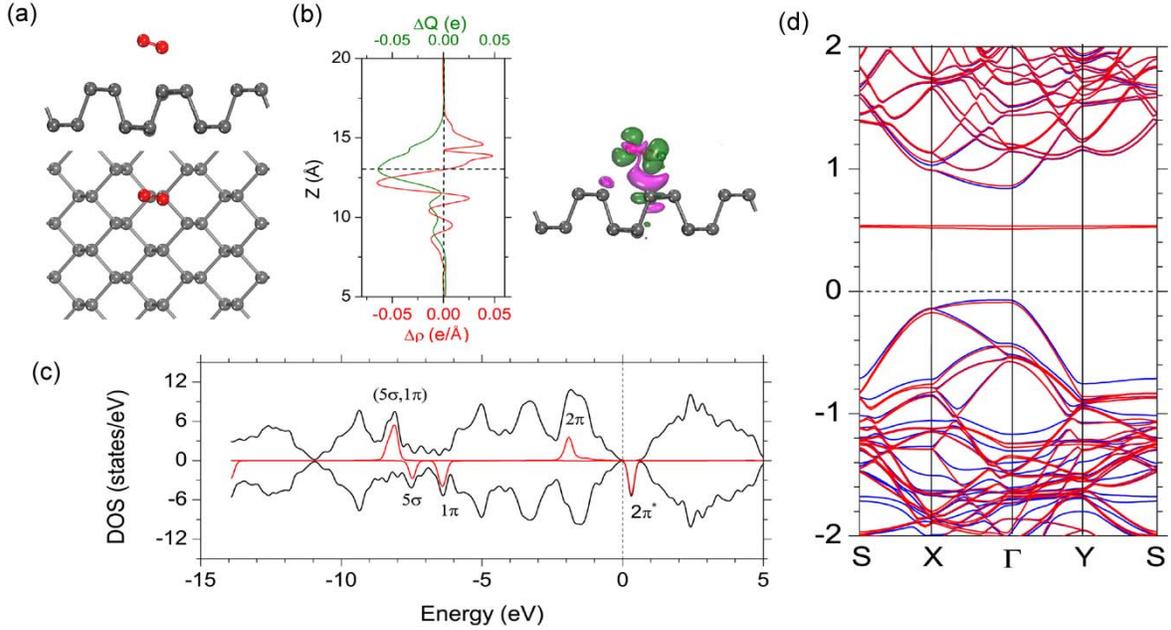

Fig. 7 $O_2$ adsorbed on phosphorene. (a) The side and top views of the lowest-energy configuration. The phosphorus and oxygen atoms are represented by balls in grey and red colors, respectively. (b) Plane-averaged differential charge density $\Delta\rho(z)$ (red line), amount of transferred charge $\Delta Q(z)$ (green line), and side view of the 0.005 Å$^{-3}$ DCD isosurface (right). The violet (green) color denotes diminishing (accumulation) of electrons. (c) DOS of $O_2$ on phosphorene (black line) and LDOS and positions of the molecular orbitals of $O_2$ (red line). (d) Spin-polarized band structures of $O_2$ on phosphorene with spin-up bands and spin-down bands are indicated in blue and red, respectively.

Table 1. The binding energy ($E_b$), the charge transfer ($\Delta q$) from molecule to phosphorene, and the X-P bond length ($B_{X-P}$), where X represents the atom in the molecule. A positive $\Delta q$ indicates a transfer of electrons from the molecule to phosphorene. We report only the range of lengths where there are several X-P bonds. The oxidation state of the molecule on phosphorene is also listed, and compared with its adsorption on graphene according to Ref. 16.

| Molecules | $E_b$(eV) | $\Delta q$(e) | $B_{X-P}$ (Å) | Molecule on Phosphorene | Molecule on Graphene |
|---|---|---|---|---|---|
| CO | -0.31 | 0.007 | 3.06-3.15 | Donor | Donor |
| $H_2$ | -0.13 | 0.013 | 2.46 | Donor | |
| $H_2O$ | -0.14 | 0.035 | 2.71 | Donor | Acceptor |
| $NH_3$ | -0.18 | 0.050 | 2.59 | Donor | Donor |
| NO | -0.32 | -0.074 | 2.75-2.88 | Acceptor | Donor |
| $NO_2$ | -0.50 | -0.185 | 2.27 | Acceptor | Acceptor |
| $O_2$ | -0.27 | -0.064 | 2.76 | Acceptor | |

## DISCUSSION

Owing to its high mobility, phophorene is promising for fabricating nanoelectronics devices complementary to graphene. However, phosphorene tends to suffer from the disturbance of chemical interactions from its surrounding molecules. Here, our work shows that phosphorene is electronically polarized upon the physisorption of many small gas molecules. Such a high sensitivity of phosphorene to those physisorbed small molecules renders it promising for gas sensor applications. The marked charge transfer upon exposing phosphorene to $H_2O$, $NH_3$, NO, $NO_2$ and $O_2$ molecules suggests that molecular physisorption is an effective approach to modulate the carrier density of phosphorene, similar to chemical doping used in a field-effect transistor. Because of its atomically thin thickness, an even moderate charge transfer between phosphorene and the adsorbed molecules can markedly increase its charge carrier density. Previous experiment on $MoS_2$ has shown that molecular gating can conveniently modulate the carrier density much beyond the dielectric breakdown point.[14]

It should be noted that while in the present study, we only considered dilute doping by putting a single molecule on the large supercell. In high doping content, the adsorption energy and the magnitude of charge transfer of each molecule may be different from the prediction here due to the repulsive electrostatic interaction between the adsorbed molecules.[53] We have performed additional calculations by considering two more types of supercell (2×3 and 5×4). Adsorptions of NO and $NO_2$ are chosen as representatives to identify the lowest-energy configuration among several potential adsorption geometries to consider the size effect. Our results show that the change in supercell size causes a change in absorption configuration for NO molecule, while the absorption configuration for $NO_2$ molecule seems to be not affected by this change. Since the change in supercell size changes the absorption density, this indicates that the content of the gas can affect the absorption configuration. Interestingly, the oxidation state ("acceptor") for both molecules is unchanged and thus seems to be independent of the adsorption sites and geometries. Since phos-phorene displays a large thickness dependence of work function and band gap.[33] future work on the interaction of small molecules physisorbed on multilayer phosphorene is highly desired.

Charge transfer between external molecules and graphene was found to alter the strength of C-C bonds through identifying the shift of Raman G-band.[54,55] It is expected the charge transfer between external molecules and phosphorene can also alter the strength of P-P bonds. In general, donor molecules are able to strengthen the P-P bonds, whereas acceptor molecules are able to weaken the P-P bonds. Hence, future experimental measurements of the shifts of Raman or IR peaks of the phonon modes is highly desired as the softening or hardening shifts of these frequencies can provide important information for understanding the interaction between external molecules and phosphorene. In addition, we find that all the open-shell molecules ($NO$, $NO_2$, and $O_2$) tend to have a larger binding energy than the closed-shell counterparts, and the adsorbed phosphorenes have new in-gap states induced by the open-shell molecules, which may create new recombination centers for excitons and trigger some new effects on the optical properties of phosphorene.

## CONCLUSION

The noncovalent interaction between a number of small molecules and phosphorene is investigated by detailed first-principles calculations. We demonstrate that the physisorbed molecules can significantly alter the electronic properties of phosphorene. $NO_2$ and NO act as a strong acceptor while $NH_3$ and $H_2O$ serve as a donor to phosphorene. Owing to the differences in work function between phosphorene and graphene and in the molecular orbital alignment, both NO and $H_2O$ molecules play a distinctively different role in doping phosphorene compared with graphene: NO accepts and $H_2O$ donates electrons upon their adsorption on phosphorene; while NO is a donor and $H_2O$ is an acceptor on graphene. The marked modification in the electronic and magnetic properties of phosphorene induced by the open-shell molecules like $NO_2$, NO, and $O_2$, together with a strong binding energy, suggests that phosphorene is promising for molecular sensor applications. The moderate adsorption energy of $H_2$ may be an advantage for phosphorene to serve as an effective hydrogen storage material at room temperature.

## AUTHOR INFORMATION

**Corresponding Author**

zhangg@ihpc.a-star.edu.sg; zhangyw@ihpc.a-star.edu.sg

**Notes**

The authors declare no competing financial interests.

## ACKNOWLEDGMENT

The authors gratefully acknowledge the financial support from the Agency for Science, Technology and Research (A*STAR), Singapore and the use of computing resources at the A*STAR Computational Resource Centre, Singapore.

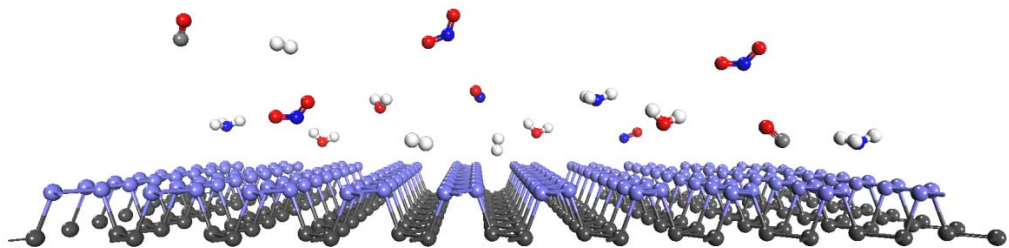